# Unraveling longitudinal field mediated versatile Stokes polarimetry

Shuaijie Yuan[1†], Xu Zhu[1†], Jin Yang[1†], Yu Liu[1], Jinhai Zou[1], Zhongquan Nie[1*], Baohua Jia[2*] and Bing Lei[1*],

[1]*College of Advanced Interdisciplinary Studies, National University of Defense Technology, Changsha, 410073, China*

[2]*Centre for Atomaterials and Nanomanufacturing (CAN), School of Science, RMIT University, Melbourne, 3000 VIC, Australia*

[†]*These authors contributed equally to this work.*

*Correspondence: ZQ Nie, E-mail: niezhongquan1018@163.com; B Lei, email: leibing_2000@nudt.edu.cn; BH Jia, E-mail: baohua.jia@rmit.edu.au*

**Abstract:**

Stokes polarimetry has been considered as an alluring platform that enables a plethora of applications ranging from single-molecule orientation to deep-space sensing. Existing polarimetry avenues, however, rely primarily on the transversely polarized field reconstruction, thus suffering from several challenges such as multiple time-sequenced detections, complex demodulation algorithms, and intricate engineering procedures. To circumvent these challenges, here we first demonstrate a longitudinally polarized field mediated recipe for the realization of efficacious and refined Stokes polarimetry in situ. This is achieved by unraveling the spin-to-orbit momentum conversion under non-paraxial focusing conditions enabling the direct mapping of the polarization ellipse. Leveraging this mechanism, we reveal an analytical solution of polarization ellipse via the local sampling of longitudinal field, in which the robustness can be fairly reinforced by relevant global retrieval based on convolutional neural network. The resultant Stokes polarimetry is shown to simultaneously exhibit in-situ signal acquisition (direct discernment), unparalleled demodulation time (up to microsecond level), superior detection efficiency (no need of troublesome design), and ultra-high retrieved accuracy (less than 1%), which is fundamentally inaccessible with traditional polarimetry methods. Our work holds great promise for empowering an all-in-one versatile vector polarimeter, which opens up a host of applications relevant to polarization control.

## 1. INTRODUCTION

Polarization refers to the direction in which the electric field vector of light oscillates. As a fundamental characteristic of light, it underpins a plethora of applications from biomedical diagnosis [1,2], materials science analysis [3] to photonic devices [4-7]. These applicable scenarios hinge essentially on the precise control of polarization and its efficient detection, which is a long-standing pursuit goal in Stokes polarimetry (SP) community as well. The traditional SP empowers transverse polarization reconstruction by leveraging either spatial or temporal multiplexing, thereby facing several key challenges as follows: i) to obtain the differences of orthogonal polarization field components, it requires multiple rotational operation and long time-sequenced measurement, leading to a vulnerable accuracy and poor detection efficiency; ii) to retrieve the polarization information at once, complex demodulation algorithms are essential, which are confronted with inferior demodulation speed, impeding the development of fast and high-sampling polarimeter; iii) although metasurface/metamaterial devices have revolutionized the possibility of visualizing the state of polarization directly, it is strictly limited by the intricate engineering and fabrication procedures. Therefore, there is an urgent need to develop a feasible strategy that could perform high-accurate, fast-speed, and efficient SP in situ.

In this regard, tremendous research efforts, including the transversely polarized fields' division of time and division of space, have been devoted to the development of ingenious SP configurations. The former requires rotating either polarization waveplates [8-10] or active modulators [11,12] repeatedly, resulting in a long detection time and sensitive error accumulation. This may impede some related applications, especially in real-time polarimetric sensing and imaging. As a powerful advancement, the latter has been put forward to configure multifunctional spatial multiplexing techniques. The common strategy is to spatially encode miscellaneous vectorially polarized beams by resorting to anisotropic optical elements like birefringent prisms [13-15], vortex waveplates [16-19], and gradient-index lenses [20-22], thus facilitating a huge leap in detection speed while conserving superior sensitivity in the SP community. However, all of these techniques critically share a collective limitation in that they undertake complicated demodulation algorithms to retrieve the information in the SP systems, while lacking the ability to reconstruct the polarization events directly. The closest answer to such an issue has been reported by embedding computational logic into the optical response of artificial micro-nanostructures (e.g. metasurfaces or metamaterials), enabling ultrafast single-shot polarization recognition [23-28]. Despite the remarkable achievements created by these SP devices, they suffer from twofold restrictions such as intricate engineering/fabrication processes and poor diffraction efficiency. Overall, a single SP technology with multiple functionalities is desirable for applications where single-shot, fast speed, and high accuracy/efficiency are in high demand. It is thus interesting to ask whether we can develop an innovative SP setup to perform these challenging optical tasks together, which remains to be elusive thus far.

To tackle the trilemmas described above, we first demonstrate a longitudinal-field-mediated strategy that executes multifunctional SP, in which the longitudinally polarized field serves as a pivotal knob to map the polarized ellipse directly. With this conceptual paradigm, we exploit the local sampling of longitudinal field to yield the polarization visualization. Notably, the global CNN optimization is further invoked to suppress the experimental pixel-level jitters of polarization retrieval. As a result, we succeed in achieving all-in-one SP with high reconstruction accuracy, fast modulation speed, and favorable detection efficiency-all of these traits that would require the mutual synergy of various techniques if implemented otherwise. The underlying principle originates from the asymmetric energy transfer from the transverse components to the longitudinal counterpart under the non-paraxial focusing condition due to the microscopic spin-orbit coupling of light. We envision this advanced SP platform to inspire new directions in dynamic polarization control, customized light-matter interaction, and real-time three-dimensional sensing.

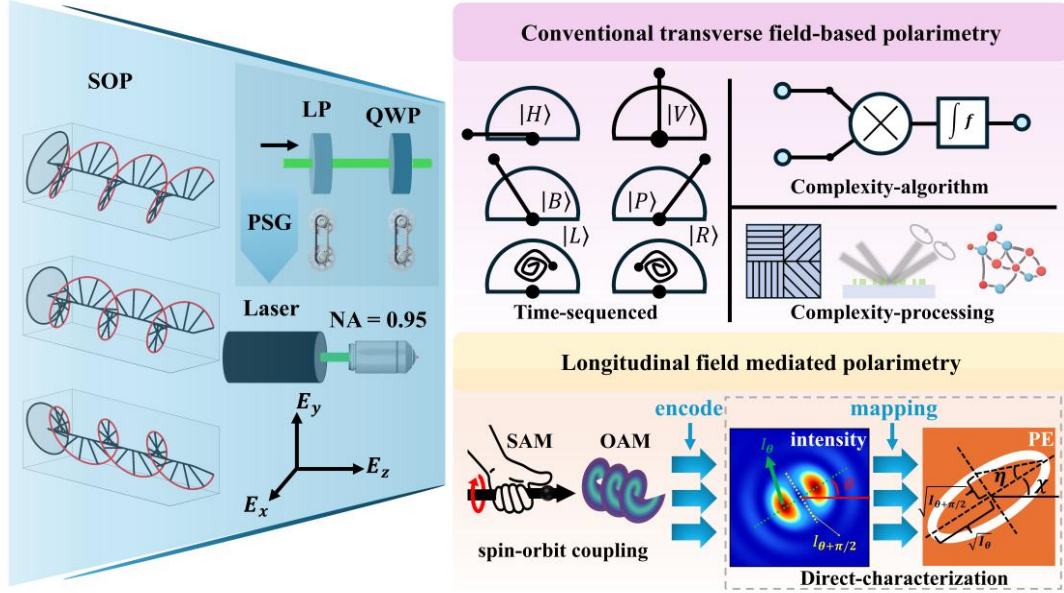

**Fig. 1. The concept and mechanism of longitudinal field mediated Stokes polarimetry. (a)** Schematic of arbitrary polarization state generation and tight focusing. A polarization state generator (PSG) consists of a rotating linear polarizer (LP, controlling the azimuth angle) and a quarter-wave plate (QWP, controlling the ellipticity angle), which is then focused by a high numerical aperture objective (NA=0.95) to produce three orthogonal electric fields ($Ex, Ey, Ez$). **(b)** Schematics of conventional Stokes polarimetry based on transverse fields. The time-division multiplexing scheme (left) using a set of orthogonal polarization bases, such as horizontal ($|H\rangle$), vertical ($|V\rangle$), ±45° linear ($|P\rangle$, $|B\rangle$), and circular ($|L\rangle$, $|R\rangle$) states, leading to a time-sequencing limitation. The spatial-division multiplexing scheme (right) enables single-shot measurement but introduces complexities in algorithm demodulation and device processing. **(c)** Schematic of the proposed longitudinal field mediated Stokes polarimetry. Under non-paraxial focusing, spin-orbit coupling (SOC) encodes the incident polarization information into the intensity distribution of the longitudinal field ($Ez$) via a helical phase. This resulting intensity profile directly maps to the polarization ellipse (PE, as indicated by $\theta = \chi$ and $\eta = \tan\sqrt{\frac{I_{\theta+\pi/2}}{I_\theta}}$) without complicated design.

## 2. RESULTS

**Design concept.** Our objective is to enact an on-demand field for mapping polarization ellipse

(PE) visually. Unfortunately, the reconstruction of PE dictated by transverse fields suffers from considerable dilemmas. It is therefore desired to seek a novel avenue for this problem. In essence, we expect that the proposed method to not only deliver anisotropic-element-like spatial encoding without complex algorithm but also offer metasurface-like direct SOP visualization without cumbersome fabrication. According to these well-defined ideas, we recall that—under non-paraxial conditions—the longitudinal fields become non-negligible relative to the transverse fields due to the depolarization effect. In this configuration, a longitudinal vortex field can be produced when the incident field is elliptically polarized. It is thus interesting to ask whether the longitudinal field distribution rigidly matches with the signatures of PE (that is, azimuth and ellipticity angles) when the spin angular momentum (SAM, elliptical polarization) converts to the orbit angular momentum (OAM, vortex phase). The answer is "yes".

We next will derive the generation of longitudinal vortex fields by strongly focusing the elliptically polarized light. Figure 1 outlines the proposed methodology. The SOP can be described by the PE, which is equivalently represented by the electric field vector trajectory, as:

$$\begin{bmatrix} E_{0x} \\ E_{0y} \end{bmatrix} = E_0 \begin{bmatrix} \cos\chi & -\sin\chi \\ \sin\chi & \cos\chi \end{bmatrix} \begin{bmatrix} \cos\eta & i\sin\chi \\ i\sin\eta & \cos\eta \end{bmatrix} \begin{bmatrix} 1 \\ 0 \end{bmatrix} = E_0 \begin{bmatrix} \cos\chi\cos\eta - i\sin\chi\sin\eta \\ \sin\chi\cos\eta + i\cos\chi\sin\eta \end{bmatrix}. \quad (1)$$

As illustrated in Fig. 1a, the SOP is experimentally modulated using a linear polarizer (to control the ellipticity angle $\chi$) and a quarter-wave plate (to control the azimuth angle $\eta$). Typically, determining the SOP on the Stokes-Poincaré sphere requires a minimum of four linearly independent measurements. In practice, six or more detection channels are often employed to enhance noise immunity and accuracy. Conventional polarimetry techniques, while effective, are inherently constrained by their reliance on transverse field detection mechanisms, as depicted in Fig. 1b. This requirement for either multiple measurements or intricate processes including modulation (in device design and fabrication) and demodulation (in algorithm), which causes accuracy/efficiency loss. An opportunity arises, however, the pronounced wavefront bending couples a portion of the transverse field components into the longitudinal direction. According to the vectorial projection relationship, the electric field at the exit pupil $\vec{E}_t$ can be expressed as:

$$\vec{E}_t = \vec{E}_{inc} - (\vec{s} \cdot \vec{E}_{inc})\vec{s}. \tag{2}$$

Here, the wave vector is $\vec{s} = (s_x, s_y, s_z) = (\sinθ\cosφ, \sinθ\sinφ, \cosθ)$, and the incident field at the entrance pupil is $E_{inc} = (E_{0x}, E_{0y}, 0)$. According to the Richards-Wolf vector diffraction theory, the focal field, formed by the coherent superposition of wave contributions from all points on the spherical exit pupil wavefront, can be written as:

$$E(r) = -\frac{ikf}{2\pi} \int_0^{\theta_{max}} \int_0^{2\pi} \sqrt{\cos\theta}\, E_t e^{(ik\vec{s}\cdot\vec{r})} \sin\theta d\phi d\theta, \tag{3}$$

where $\vec{r} = (\rho\cos\varphi, \rho\sin\varphi, z)$ is the cylindrical coordinate vector and $\sqrt{\cos\theta}$ represents the apodization factor. Substitution of the PE (Eq. 1), simplifies to the analytical expression under focal plane:

$$E_z(\rho,\varphi) = -kf I(\rho) E_0 [-\frac{\sqrt{2}}{2}\sin\left(\eta - \frac{\pi}{4}\right)e^{i(\chi-\varphi)} + \frac{\sqrt{2}}{2}\cos\left(\eta - \frac{\pi}{4}\right)e^{-i(\chi-\varphi)}], \tag{4}$$

the integral kernel $I(\rho) = \int_0^{\theta_{max}} \sqrt{\cos\theta}\sin^2\theta\cos\theta J_1(k\rho\sin\theta)d\theta$ depends solely on the radial distance $\rho$, and $J_1(\cdot)$ is the first-order Bessel function. Eq. 4 reveals that the phase of $E_z$ is jointly modulated by a pair of positive and negative vortex phases ($e^{\pm i\varphi}$). This phenomenon is essentially the result of spin-orbit coupling (SOC) [31,32], whereby the spin angular momentum (SAM) associated with the SOP is converted into longitudinal orbital angular momentum (OAM) during the focusing process. Specifically, the input optical field to be characterized is decomposed into left-handed and right-handed circular polarization components, which respectively excite longitudinal spatial modes carrying opposite helical phases. These modes superpose coherently at the focal point, yielding a symmetry-broken annular intensity distribution:

$$I_z = |E_z|^2 \propto [cos^2\eta cos^2(\chi - \varphi) + sin^2\eta sin^2(\chi - \varphi)]. \tag{5}$$

$$I_z(\varphi = \chi) \propto cos^2\eta \text{ and } I_z(\varphi = \chi + 90°) \propto sin^2\eta. \tag{6}$$

This leads to a concise relationship: the azimuth angle $\chi$ corresponds to the inclination angle $\theta$ (from the physical coordinate of the maximum intensity, as can be verified by differentiating $I_z$, see Section S1 in Supplementary Materials). Furthermore, the ellipticity angle $\eta$ is determined by the intensity values along two orthogonal axes, specifically, $I_z(\varphi = \chi)$ and $I_z(\varphi = \chi + 90°)$, corresponding to the major and minor axes of the polarization ellipse, respectively.

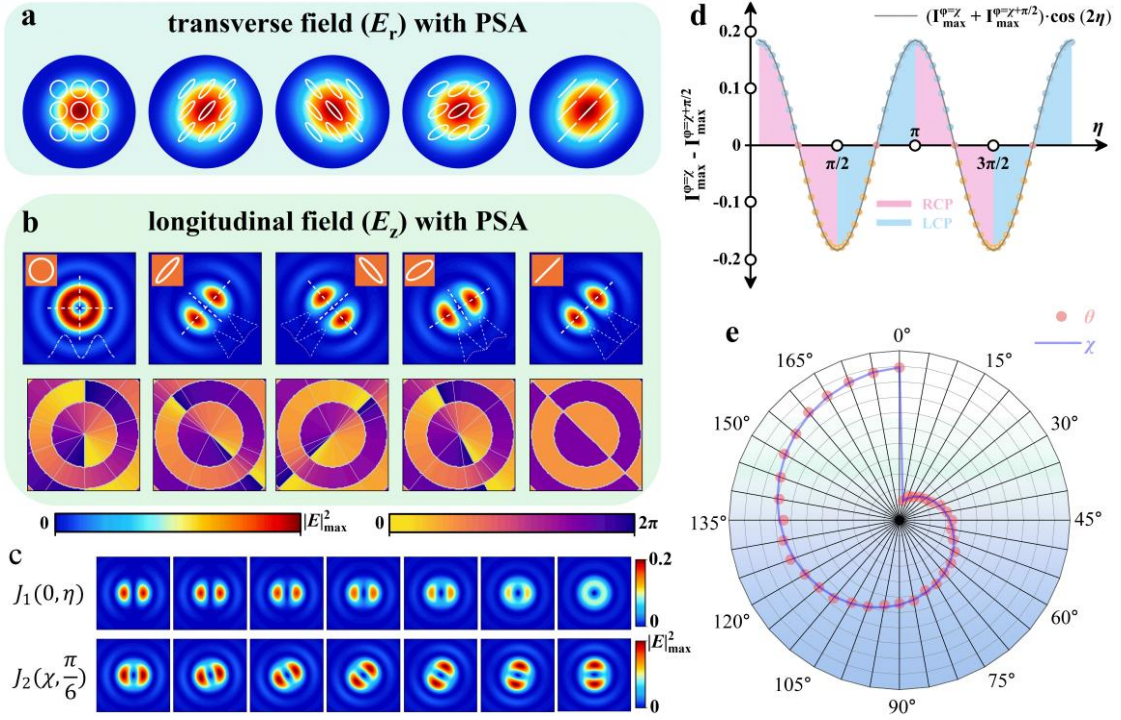

**Fig. 2. Theoretical analysis on the local sampling of longitudinal field mediated Stokes polarimetry.** (a) Schematic of conventional polarization detection based on transverse fields, which relies on a complex polarization state analyzer (PSA). (b) Intensity and phase distributions of the longitudinal field component for different input SOPs (depicted in the orange-framed insets). (c) Evolution of the longitudinal field intensity distribution when varying the ellipticity angle (e.g., $J_1(0,\eta)$, varied in steps of $\pi/24$) at a fixed azimuth angle, and the azimuth angle (e.g., $J_2(\chi, \pi/6)$, varied in steps of $\pi/12$) at a fixed ellipticity angle. (d) Functional relationship between the ellipticity angle $\eta$ and the maximum intensity along two orthogonal axes under a fixed azimuth angle $\chi$. (e) Correspondence between the azimuth angle $\chi$ and the inclination angle $\theta$ of the intensity pattern at a fixed ellipticity angle $\eta$.

Based on the above theoretical framework, the mapping rule derived from the analytical solution is verified against simulations in Figure 2. For comparison, Fig. 2a shows that the transverse field intensity offers only a negligible response to variations in the input SOP after strong focusing, thus necessitating a roundabout and complex PSA for detection. In distinct contrast, Fig. 2b reveals a direct isomorphism between the longitudinal field distribution and the PE shape. This correspondence manifests as a symmetry-broken annular intensity profile that is unique for each SOP. These distinct intensity profile originates from the deterministic conversion of the input SOP, which carries a specific SAM, into a corresponding longitudinal OAM under tight focusing [31, 32]. The resulting OAM state is characterized by a helical wavefront with a spatially varying phase gradient. It is this specific gradient distribution that sculpts the final asymmetric intensity pattern, establishing a one-to-one correspondence between the input SOP and its focal signature.

Interestingly, Fig. 2c reveals that the influences of azimuth and ellipticity angles on the longitudinal field distribution are mutually independent. On the one hand, when the azimuth angle $\chi$ fixed at 0° and the ellipticity angle $\eta$ is incremented in steps of $\pi/24$ (i.e., $J_1(0,\eta)$), the intensity profile evolves via an energy redistribution between two orthogonal axes. This process obeys energy conservation, with energy transferring from the horizontal to the vertical axis, until a new equilibrium is reached. Phase analysis indicates that the orientation of the phase interfaces (0 and $2\pi$) remains fixed, while their curvature varies continuously (see the phase density and circular-line-scan profiles in Fig. S1, Supplementary Material). On the other hand, when the ellipticity angle $\eta$ is fixed at $\pi/6$ and the azimuth angle $\chi$ is scanned in steps of $\pi/12$ (i.e., $J_2(\chi,\pi/6)$), the intensity pattern rotates around the origin without axial energy transfer. Here, the phase gradient maintains a constant curvature, but the interface orientation rotates (see the phase density and circular-line-scan profiles in Fig. S1, Supplementary Material).

Based on this phenomenon, to verify the mapping rules derived from the theoretical analytical solution, we numerically calculated the variations of $J_1(0,\eta)$ and $J_2(\chi,\pi/6)$ with respect to the ellipticity and azimuth angles, respectively. Because the domain of the square root operation constrains the ellipticity angle $\eta$ to positive values, a minor mathematical modification is applied (see Supplementary Eqs. S30-S35). As shown in Fig. 2d, the cosine term $\cos(2\eta)$ excellently matches the varying intensity difference on the orthogonal axes ($I_z^{\varphi=\chi} - I_z^{\varphi=\chi+\pi/2}$) when normalized by a constant ($I_z^{\varphi=\chi} + I_z^{\varphi=\chi+\pi/2}$), demonstrating excellent consistency with the theory. Similarly, we defined the inclination angle $\theta$ as the angle between the horizontal axis and the line connecting the origin to the peak-intensity direction. The radar chart in Fig. 2e shows that the inclination angle matches the azimuth angle exactly (i.e., $\chi = \theta$). A slight deviation is observed, attributable to the fact that pixels in the numerical simulation represent discrete grid points rather than infinitesimal physical points and a detailed analysis is provided in Supplementary Material Section S5.

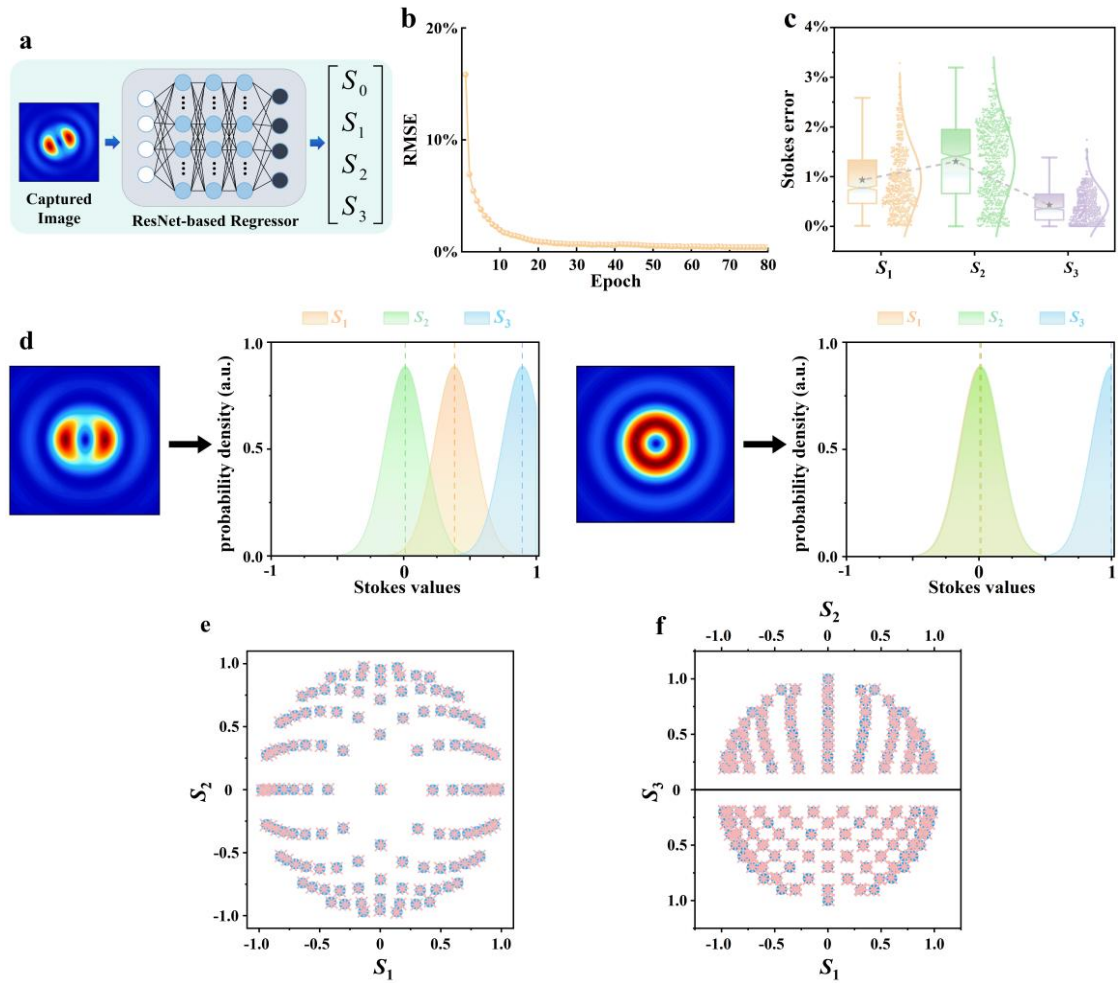

**Fig. 3. Diret prediction on global sampling of AI-assisted longitudinal-field-mediated Stokes polarimetry. (a)** ResNet-34 framework for predicting Stokes parameters from longitudinal field images. **(b)** The decline curve of root means square error (RMSE) with the training rounds. **(c)** Stokes parameter error distribution from 1000 samples. **(d1-d2)** CNN-derived probability distributions of Stokes parameters, with peaks (predictions) marked against true values. **(e, f)** Poincaré sphere projections comparing theoretical (circles) and predicted (crosses) values.

The above theoretical framework and numerical simulations establish a direct correspondence between the SOP and the longitudinal field distribution. The analytical solution derived from this correspondence can be obtained through local sampling in only 0.45 ms (i7-14650HX, 20-run average, operations: two square roots, one division, one tangent). However, this ideal mapping is often disrupted by experimental imperfections such as beam jitter, speckle flickering, intensity drift, optical aberrations, and detector noise. These factors introduce localized intensity anomalies that impair the robustness of purely analytical computations. To address these challenges, we employ a convolutional neural network (CNN) to learn the Stokes parameters directly from the global patterns in the intensity images. This approach enhances robustness by leveraging holistic image modeling to suppress local anomalies and compensate for optical imperfections, eliminating the need for specialized error-specific correction algorithms. We utilize ResNet-34 as the backbone network (Fig. 3a) to balance deep feature extraction with computational efficiency, creating an end-to-end pipeline from raw image input to Stokes vector output. The network is adapted for single-channel intensity images by modifying the input layer. The final classification layer is replaced with a regression head comprising fully connected layers with Batch Normalization and ReLU activation, supplemented by dropout to prevent overfitting. The output layer consists of three neurons with Tanh activation, constraining the predictions to the physically valid range of $[-1, 1]$. The model is trained on a comprehensive simulated dataset of 10,000 images covering a wide distribution of Stokes parameters, enabling it to capture the underlying relationship between global intensity patterns and polarization states. To respect the geometric properties of the Stokes vector on the Poincaré sphere, we designed a custom loss function based on the quaternion representation of the rotation between the predicted and true Stokes vectors. This method, which respects the spherical topology of the Poincaré sphere, effectively avoids the gradient explosion problems that can occur at the domain boundaries with other approaches. After 80 training epochs, the RMSE converges to below 1% (Fig. 3b), indicating stable and effective model performance.

To evaluate the generalization capability of CNN system, we examine the inference error distributions for the three Stokes parameters across 1000 test samples, as displayed in Fig. 3c. The results show that most errors (>90%) are below 2%, and the low variance indicates high stability. Notably, although some samples may exhibit a relatively high error in one parameter, the errors in the other parameters are significantly lower. As a result, the overall error remains below 1%. This error level is practically insignificant, given the accuracy limits of commercial polarimeters. These results validate the ability to accurately extract Stokes parameters from the overall image features. Figs. 3d1-d2 visualize the decision-making mechanism by showing the distribution of the network's predictions for two sample images. The sharp peaks align closely with the true values. For these samples, the network directly regresses the Stokes parameters, achieving remarkably low prediction errors of 0.71% and 0.63%, respectively. To further quantify the performance, we compare 121 right-handed elliptically polarized states uniformly distributed on the Poincaré sphere through three orthogonal projections (Figs. 3e-3f), observing strong agreement between theoretical and predicted values. Unlike analytical methods that rely on local features, the integrated CNN effectively mitigates inevitable non-ideal effects encountered in practical experiments. This demonstrates that the data-driven approach employed by the CNN offers enhanced robustness and is better suited for complex environments.

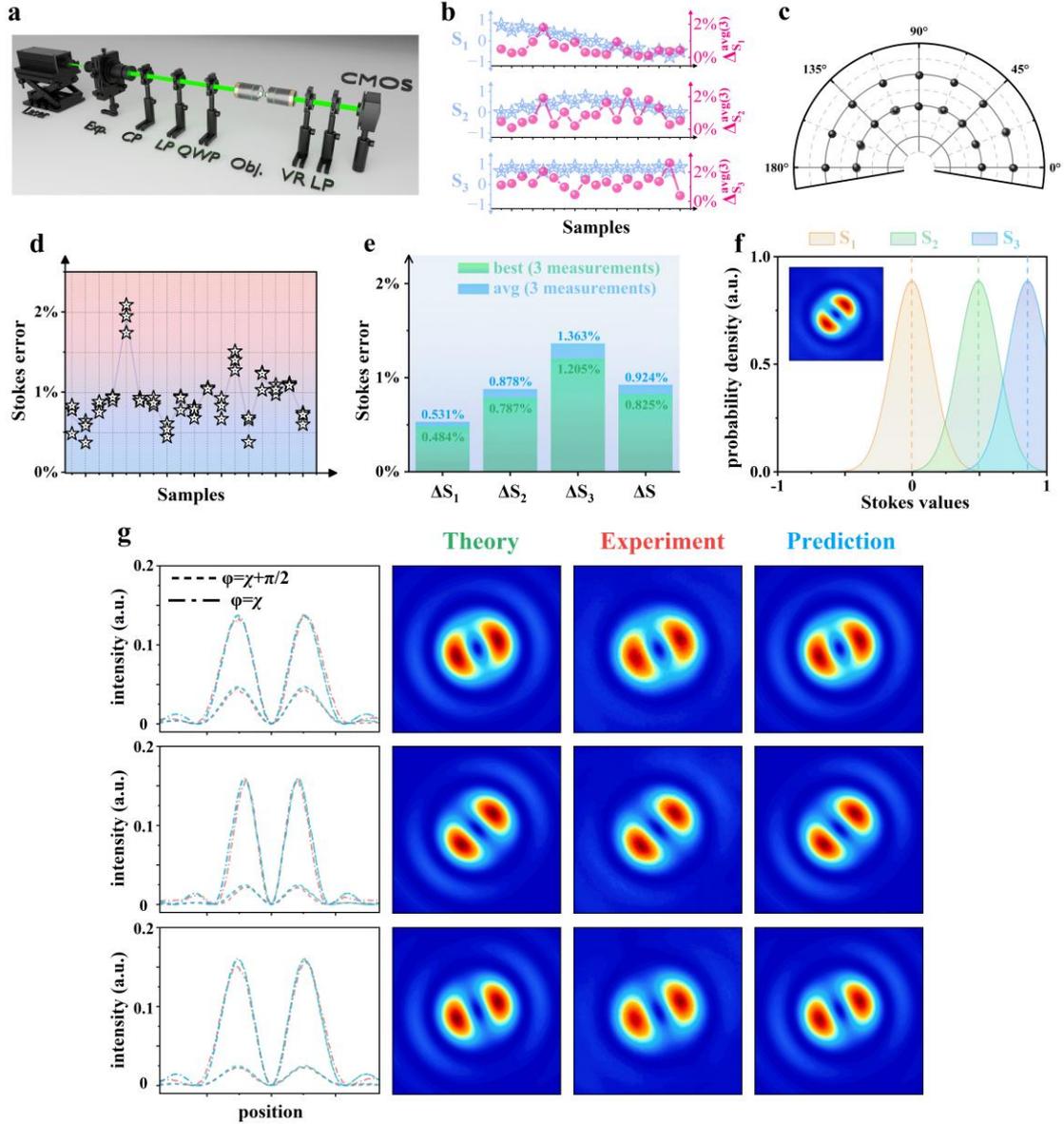

**Fig. 4. Experimental verification of longitudinal-field-mediated Stokes polarimetry. (a)** Experimental configuration. CP, circular polarizer; LP, linear polarizer; QWP, quarter-wave plate; Exp., beam expander; obj, high-NA objective lens (NA=0.95); VR, Vortex retarder. **(b)** Normalized Stokes parameters and their errors for three measurements across 18 SOPs. Stars and blue dots denote predicted and true values, respectively (left axis); spheres represent errors (right axis). **(c)** Top-view projection of the Poincaré sphere showing 54 predictions. True values span azimuth angles from 0° to 90° in 11.25° increments, with ellipticity angles of 20.7° and 30°. **(d)** Error distribution for the 54 predictions, with the light purple curve representing the average error distribution. **(e)** The optimal and average errors across three measurements,

averaged over all 18 SOPs. **(f)** CNN-derived probability density distributions for the predicted Stokes parameters of a representative sample. **(g)** Comparison of numerically simulated, experimentally captured, and predicted (reconstructed from predictions) intensity images and their profiles along two orthogonal axes.

Building upon the confirmed robustness of the holistic CNN approach, its performance was experimentally validated under practical conditions using the single-shot polarimetry setup illustrated in Fig. 4a. The system incorporated a polarization state generator comprising a polarizer and a quarter-wave plate. After strong focusing, the resulting longitudinal field is analyzed by a detector module containing a vortex retarder and a polarizer [32]. It is noteworthy that even under standard operation, repeated measurements of the same SOP exhibited noise, artifacts, and intra-class variability. To rigorously assess the method's ability to handle these non-ideal effects, we selected 18 distinct SOPs and acquired three images for each at varying intervals, introducing sampling variations. As presented in Fig. 4b, the Stokes parameters and errors derived from the 54 samples show close agreement between ground-truth values (blue dots) and predictions (stars) on the left axis, while the right axis displays errors relative to the triplicate measurement mean, yielding an overall average error of approximately 1%. The tight clustering of all predicted points on a Poincaré sphere projection in Fig. 4c further confirms robustness against local fluctuations. The error distribution across samples in Fig. 4d indicates that over 70% of errors are below 1%, with consistently low variation per SOP. Quantitative results in Fig. 4e list the average errors for the three measurement sets as 0.531%, 0.878%, and 1.363% (overall mean: 0.924%), with the corresponding minimum errors being 0.484%, 0.787%, and 1.205% (overall mean: 0.825%)—all the overall mean values remain below 1%. For a representative sample, the regression output in Fig. 4f shows a sharp probability peak aligned with the true value, yielding a minimal error of 0.073%. Fig. 4g compares the simulated, experimentally captured, and predicted field images, with the intensity profiles along the inclination angle axis and its orthogonal direction showing excellent consistency. Compared to existing methods, our approach simultaneously achieves high accuracy, rapid response, simple structure, and strong robustness, representing a significant advance for Stokes polarimetry.

## 3. CONCLUSIONS AND DISCUSSIONS

To conclude, we have proposed and experimentally demonstrated a versatile SP paradigm via wavelength-scale longitudinal field, which originates from the native transformation from the transverse SAM to longitudinal OAM of elliptically polarized light upon tight focusing. Based on this new concept, we are able to garner the analytical solution of longitudinal field to locally and directly map the PE (i.e. azimuth and axis length). Deep learning is further exploited to reinforce the global polarization extraction, thus enabling unprecedented polarization information acquisition. More impressively, the proof-of-concept experiments unravel that high-accuracy (<1%), excellent-efficiency (without complicated design), and fast-modulation (~450 μs on an i7-14650HX with extremely low computing resource consumption) SP can be achieved, which is in a good agreement with the above theoretical elaboration and neural network prediction. Notably, our longitudinal-field-mediated-SP solution is free from rotated polarization components, complex demodulation algorithms, cumbersome micro-nano processing, and tedious optical path engineering, which are extremely challenging or impossible with conventional SP methods. Our advanced SP has proven applicability and holds promise for real-time polarization sensing, polarization control and optical quantum communication.

Overall, our work establishes a bran-new versatile longitudinal field mediated SP platform, which would provide us with some privileges in the foreseeable future. First, it offers a feasible essential route to customize higher-dimensional polarization field control, which may allow us to achieve ultra-precise, high-speed, and nanometer-scale SP. Second, it sheds light on the exploration of advanced light-matter interaction, not only making the realization of exotic full Mueller matrix polarimetry possible, but also facilitating the development of nonlinear SP. Third, several applications would benefit from such a distinct SP paradigm. On the one hand, the sub-wavelength characteristic in the polarization detection enables seamless integration with commercial optical microscopy. This is adapted to the super-resolved polarimetric imaging for live-cell observation and non-invasive tissue diagnosis. In parallel, it can also be utilized to develop fast-speed and high-efficiency poalrimetric detection/sensing techniques where the longitudinal field plays a key role. Besides, it provides a powerful toolkit to permit large-scale polarimetric datasets, thus favoring to train neural networks for machine-learning-assisted classification in polarization-related applications.

**Acknowledgments.** Z. Nie acknowledges the support partly from the *National Natural Science Foundation of China* (Grants Nos. 12574333 and 11974258) and *Innovation Science Fund of National University of Defense Technology* (Grant No. 25-ZZCX-JDZ-28). B. Lei acknowledges the support partly from the *National Science Foundation of China* (Grants Nos. 62475285, 61975235). J. Zhou acknowledges the support partly from the *National Natural Science Foundation of China* (Grants Nos. 62575305 and 62305275) and *Youth Independent Innovation Science Foundation of the National University of Defense Technology* (Grant No. ZK24-20). S. Yuan acknowledges the support partly from the *Hunan graduate Innovation project* (CX20240112). X. Zhu acknowledges the support partly from the *Hunan graduate Innovation project* (XJQY2024026).

**Disclosures.** The authors declare no competing interests.

**Data availability.** The data that support the plots within this paper and other findings of this study are available within this article and its Supplementary Information file are also available from the corresponding author upon request.

**Supplemental document.** See Supplement for supporting content.

**Author contributions.** S. Yuan and Z. Nie conceived this idea and conducted theoretical research. X. Zhu designed and optimized neural networks. S. Yuan conducted the experiment and analyzed the data with assistance from J. Yang Y. Liu Meanwhile, all the authors have contributed to the writing of this paper. Z. Nie, B. Jia and B. Lei supervised this project.